\documentclass[amssymb,prl,twocolumn,showpacs]{revtex4}
\input{psfig.sty}
\usepackage{epsfig}
\usepackage{dcolumn}
\usepackage{amsmath}
\begin{document}
\title{\bf AC-driven double quantum dots as spin pumps and spin filters.}
\author{Ernesto Cota$^1$, Ram\'on Aguado$^2$ and Gloria Platero$^2$}
\affiliation{1-Centro de Ciencias de la Materia Condensada -
UNAM, Ensenada, Mexico.\\ 2-Instituto de Ciencia de Materiales,
CSIC, Cantoblanco, Madrid,28049, Spain.}
\date{\today}
\begin{abstract}
We propose and analyze a new scheme of realizing {\it both} spin
filtering and spin pumping by using ac-driven double quantum dots
in the Coulomb blockade regime. By calculating the current through
the system
in the sequential tunneling regime,
we demonstrate that the spin polarization of the current can be
controlled by tuning the parameters (amplitude and frequency) of
the ac field.
We also discuss spin relaxation and decoherence effects in the
pumped current.
\end{abstract}
 \maketitle
The emerging field of spintronics aims at creating devices based
on the spin of electrons \cite{spintronics}. One of the most
important requirements for any spin-based electronics is the
ability to generate a spin current. Proposals for generating
spin-polarized currents include spin injection by using
ferromagnetic metals \cite{ferro} or magnetic semiconductors
\cite{semimag}. Alternatively, one may use quantum dots (QDs) as
spin filters or spin pumps \cite{spin-pump1,spin-pump2}.
\begin{figure}
\epsfig{file=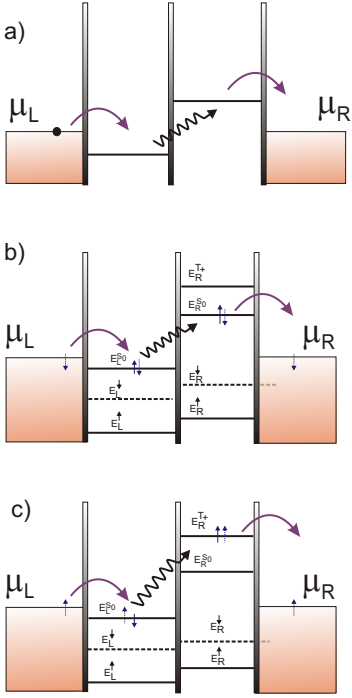,angle=0,width=0.45\textwidth}
\caption{(Color online) Schematic diagram of the double dot
electron pump. a) Pumping through one-particle states: the current
is spin-unpolarized. b) Pumping through two particle states with
$\hbar\omega_\downarrow=E_{R}^{S_0}-E_{L}^{S_0}$ (where
$E_{L}^{S_0}$ and $E_{R}^{S_0}$ are the energies of the singlets).
If the chemical potentials fulfil the conditions
$\mu_{1,0}(2,1)<\mu_L$, $\mu_{2,0}(1,2)>\mu_R$ and
$\mu_{2,1}(1,2)<\mu_R$ (see text) the resulting pumped current is
spin-down polarized. c) Pumping
involving a triplet ($E_{R}^{T_+}$)in the right dot. In this case
the pumped current is spin-up polarized. Note that during the
pumping process only electrons with spin down (case b) or up (case
c) become delocalized by the microwaves. Dashed arrows denote
delocalized spins whereas solid arrows denote spins that remain
localized on each dot.}
\end{figure}
For QD spin filters, dc transport through few electron states is
used to obtain spin-polarized currents
as demonstrated experimentally by Hanson {\it et al} \cite{hanson}
following the proposal of Recher {\it et al} \cite{recher}. Spin
current rectification has also been realized
\cite{Ono97-Johnson04}. The basic principle of spin pumps is, on
the other hand, closely related to that of charge pumps. In a
charge pump a dc current is generated by combining ac driving with
either absence of inversion symmetry in the device, or lack of
time-reversal symmetry in the ac signal. The range of possible
pumps includes turnstiles, adiabatic pumps or nonadiabatic pumps
based on photon-assisted tunneling (PAT) \cite{report1,report2}.
\par
In this Letter we propose and analyze a new scheme of realizing
{\it both} spin filtering and spin pumping by
using a double quantum dot (DQD), in the Coulomb blockade regime,
with time-dependent gate voltages and in the presence of a uniform
magnetic field.
The periodic variation of the gate potentials allows for a net dc
current through the device even with no dc voltage applied
\cite{stafford,hazelzet}: if the system is driven at a frequency
(or subharmonic) corresponding to the energy difference between
two time-independent eigenstates, the electrons become completely
delocalized \cite{oosterkamp,petta}. If the left reservoir
(chemical potential $\mu_L$) can donate electrons to the left dot
(at a rate $\Gamma_L$) and the right reservoir (chemical potential
$\mu_R$) can accept electrons from the right dot (at a rate
$\Gamma_R$) the system will then pump electrons from left to
right, even when there is no dc bias applied, namely $\mu_L=\mu_R$
(Fig. 1a). Starting from this pumping principle our device has two
basic characteristics: {\it i)} If the process involves
two-particle states, the pumped current can be completely
spin-polarized {\it even if the contact leads are not spin
polarized} and {\it ii)} the pumping can occur either through
singlet (Fig. 1b) or triplet (Fig. 1c) states depending on the
applied frequency, such that the {\it degree of spin polarization}
can be tuned by means of the ac field. For example, if one drives
the system (initially prepared in a state with $n=n_L+n_R=3$
electrons: $|L=\downarrow\uparrow,R=\uparrow \rangle$) at a
frequency corresponding to the energy difference between the
singlets in both dots,
the electron with spin $\downarrow$ becomes delocalized in the DQD
system. If now the chemical potential for taking $\downarrow$
($\uparrow$) electrons out of the right dot is above (below)
$\mu_R$, a spin-polarized current is generated. The above
conditions for the chemical potentials can be achieved by breaking
the spin-degeneracy through a Zeeman term $\Delta_z = g\mu_B B$,
where $B$ is the external magnetic field (which is applied
parallel to the sample in order to minimize orbital effects), $g$
is the effective g-factor and $\mu_B$ the Bohr magneton.
\par
Our main findings can be summarized in Fig. 2 where we present a
plot of the pumped current as a function of the applied frequency
for a particular choice of parameters. Importantly, the current
presents a series of peaks which are uniquely associated with a
definite spin polarization: the pumped current is 100\% spin-down
(up) polarized for $\omega=\omega_\downarrow$
($\omega=\omega_\uparrow$). The other peaks correspond to
subharmonics of the energy difference between the relevant states,
see below. This particular example illustrates how by tuning the
external ac field one can operate the driven DQD as a bipolar spin
filter with no dc voltage applied.
\par
{\it Formalism}.-- Our system consists of an asymmetric DQD
connected to two
reservoirs kept at the chemical potentials $\mu_\alpha$,
$\alpha=L,R$. Using a standard tunneling Hamiltonian approach, we
write for the full Hamiltonian
$\mathcal{H}_l+\mathcal{H}_{DQD}+\mathcal{H}_T$, where
$\mathcal{H}_l=\sum_{\alpha}\sum_{k_\alpha,\sigma}
\epsilon_{k_\alpha} c_{k_\alpha\sigma}^\dagger c_{k_\alpha\sigma}$
describes the leads and
$\mathcal{H}_{DQD}=\mathcal{H}_{QD}^L+\mathcal{H}_{QD}^R+\mathcal{H}_{L\Leftrightarrow
R}$ describes the DQD.
It is assumed that only one orbital in the left dot participates
in the spin-polarized pumping process whereas \emph{two} orbitals
in the right dot (energy separation $\Delta\epsilon$) have to be
considered. The isolated left dot is thus modelled as a one--level
Anderson impurity: $\mathcal{H}_{QD}^L=\sum_{\sigma} E_L^\sigma
d^\dagger_{L\sigma} d_{L\sigma}+ U_L n_{L\uparrow}
n_{L\downarrow}$, whereas the isolated right dot is modelled as:
$\mathcal{H}_{QD}^R = \sum_{i \sigma} E_{R i}^\sigma d^\dagger_{R
i\sigma} d_{Ri \sigma} + U_R(\sum_i n_{R i\uparrow} n_{R
i\downarrow}+\sum_{\sigma,\sigma'}n_{R 0\sigma} n_{R 1\sigma'})+ J
{\bf S}_0  {\bf S}_1$. The index $i=0,1$ denotes the two levels.
In practice, we take $E_L^\uparrow=E_{R 0}^\uparrow=0$
($E_L^\downarrow=E_{R 0}^\downarrow=\Delta_z$), so all the
asymmetry is included in the charging energies $U_R>U_L$.
Experimentally, this asymmetry can be realized by making the right
dot smaller. ${\bf S}_i = (1/2)\sum_{\sigma \sigma'} d^\dagger_{R
i\sigma} {\bf \sigma}_{\sigma \sigma'} d_{R i\sigma'}$ are the
spins of the two levels.
As a consequence of Hund's rule, the intra--dot exchange, $J$, is
ferromagnetic ($J < 0$) such that the singlet $|S_1\rangle =
(1/\sqrt{2}) (d^\dagger_{R0\uparrow} d^\dagger_{R1\downarrow} -
d^\dagger_{R0\downarrow} d^\dagger_{R1\uparrow}) |0\rangle$ is
higher in energy than the triplets $|T_+\rangle =
d^\dagger_{R0\uparrow} d^\dagger_{R1\uparrow} |0\rangle$,
$|T_0\rangle = (1/\sqrt{2}) (d^\dagger_{R0\uparrow}
d^\dagger_{R1\downarrow} + d^\dagger_{R0\downarrow}
d^\dagger_{R1\uparrow}) |0\rangle$ and $|T_-\rangle =
d^\dagger_{R0\downarrow} d^\dagger_{R1\downarrow}|0\rangle$. Due
to the Zeeman splitting $E^{T_-} > E^{T_0}
>E^{T_+} =\Delta\epsilon+U_R-J/4$. Finally, we
consider the case where $\Delta\epsilon >\Delta_z+J/4$ such that
the triplet $|T_+\rangle$ is higher in energy than the singlet
$|S_0\rangle = (1/\sqrt{2}) (d^\dagger_{R0\uparrow}
d^\dagger_{R0\downarrow} - d^\dagger_{R0\downarrow}
d^\dagger_{R0\uparrow}) |0\rangle$. $\mathcal{H}_{L\Leftrightarrow
R}=\sum_{i,\sigma}t_{LR}( d_{L\sigma}^\dagger d_{R i\sigma} +
h.c.)$ describes tunneling between dots. The tunneling between
leads and each QD is described by the perturbation
$\mathcal{H}_T=\sum_{k_L,\sigma}V_{L}(c_{k_L\sigma}^{\dag}d_{L\sigma}+{\rm
h.c.})+\sum_{i,k_R,\sigma}V_{R}(c_{k_R\sigma}^{\dag}d_{R
i\sigma}+{\rm h.c.})$. $\Gamma_{L,R}=2\pi \mathcal D_{L,R}
|V_{L,R}|^2$ are the tunneling rates. It is assumed that the
density of states in both leads $\mathcal D_{L,R}$ and the
tunneling couplings are energy-independent.
\par
We study the system by a reduced density matrix (RDM), $\rho={\rm
Tr}_L\chi$, where $\chi$ is the full density matrix, and ${\rm
Tr}_L$ is the trace over the leads. The dynamics of the RDM is
formulated in terms of the eigenstates and eigenenergies of each
isolated QD. We concentrate on the Coulomb blockade regime (with
up to two electrons per dot, which defines a basis of 20 states)
and study the sequential tunneling regime (Born-Markov
approximation). For example, the diagonal elements of the RDM read
\begin{equation}
\dot{\rho}_{ss}= - \frac{i}{\hbar}[\mathcal{H}_{L\Leftrightarrow
R},\rho]_{ss} +\sum_{m\neq s} W_{sm}\rho_{mm} - \sum_{k\neq s}
W_{ks}\rho_{ss} \label{eq-master}
\end{equation}
where $W_{ij}$ are the transition rates (calculated using a
standard Fermi Golden Rule). In addition we consider an ac field
acting on the dots, such that the single particle energy levels
become
$\epsilon_{L(R)}\rightarrow\epsilon_{L(R)}(t)=\epsilon_{L(R)}\pm\frac{eV_{AC}}{2}
\cos \omega t$,
where $eV_{AC}$ and $\omega$ are the amplitude and frequency,
respectively, of the applied field. We include spin relaxation and
decoherence phenomenologically in the corresponding elements of
the equation for the RDM. Relaxation processes are described by
the spin relaxation time
$T_1=(W_{\uparrow\downarrow}+W_{\downarrow\uparrow})^{-1}$, where
$W_{\uparrow\downarrow}$ and $W_{\downarrow\uparrow}$ are
spin-flip relaxation rates fulfilling a detailed balance. A lower
bound for the spin relaxation time $T_1$ of $50 \mu s$ with a
field $B=7.5 T$ has been obtained recently \cite{hanson2} for a
single electron in a QD using energy spectroscopy and relaxation
measurements.
In the following, we focus on zero temperature results such that
$W_{\downarrow\uparrow}=0$ and thus
$T_1=W_{\uparrow\downarrow}^{-1}$. The rate ${T_2}^{-1}$ ($T_2$ is
the spin decoherence time) describes intrinsic spin decoherence.
We take $T_2=0.1T_1$ in all the calculations
\cite{Loss-relaxation}.

In practice, we integrate numerically the dynamics of the RDM in
the chosen basis. In particular, all the results shown in the next
paragraphs are obtained by letting the system evolve from the
initial state $|\downarrow\uparrow,\uparrow\rangle$ until a
stationary state is reached. The dynamical behavior of the system
is governed by rates which depend on the electrochemical
potentials of the corresponding transitions. The electrochemical
potential $\mu_{1(2),i}(N_1 ,N_2)$ of dot $L(R)$ is defined as the
energy needed to add the $N_{1(2)}$th electron to energy level $i$
of dot $L(R)$, while having $N_{2(1)}$ electrons on dot $R(L)$
\cite{report2}.
The current from left to right is:
$I_{L\rightarrow R}(t)=\Gamma_R\sum_{s} \rho_{ss}(t)$,
with a similar expression for $I_{R\rightarrow L}$. Here, states
$|s\rangle$ are such that the right dot is occupied. For ease in
the notation, we take from now on $\hbar=e$ =1, such that
$V_{AC}$, $\omega$, etc, have units of energy.

\par
{\it Results}.--
A calculation of the stationary current, for each direction of
spin, namely $I_{tot}=\sum_{\sigma=\uparrow,\downarrow}I_\sigma$
as a function of $\omega$ (and fixed intensity $V_{AC}=V_{AC}^0=
0.7$), gives the results shown in Fig. 2.
\begin{figure}
\epsfig{file=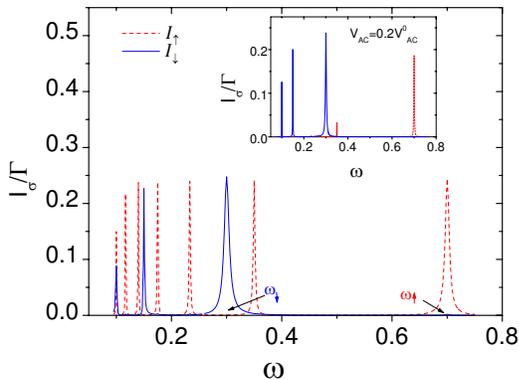,angle=0,width=0.450\textwidth} \caption{
(Color online) Pumped current as a function of the ac frequency.
The spin-down component (solid line) shows three peaks
corresponding to one ($\omega=\omega_{\downarrow}=0.3$), two
($\omega=0.15$) and three ($\omega=0.1$) photon processes,
respectively. The spin-up component (dashed line) shows a main
one-photon resonance at $\omega=\omega_{\uparrow}=0.7$ and up to
six more satellites corresponding to multiphoton processes.
Parameters: $\Gamma_{L}=\Gamma_{R}=0.001$, $t=0.005$, $U_L=1.0$,
$U_R=1.3$, $J=0.2$, $\mu_L=\mu_R=1.31$, $\Delta_z=0.026$,
$\Delta\epsilon=0.45$, $V_{AC}=V_{AC}^0=0.7$ (in meV) correspond
to typical values in GaAs QDs. In particular, the Zeeman splitting
corresponds to a magnetic field $B\approx 1T$. Inset: same as in
main figure but with a lower intensity $V_{AC}=V_{AC}^0/5=0.14$.}
\end{figure}
The main peak of $I_\downarrow$ (continuous line) occurs at
$\omega=\omega_\downarrow\equiv E_{(\uparrow,
\downarrow\uparrow)}-E_{(\downarrow\uparrow,\uparrow)}=U_R-U_L=0.3$
(see caption)
At this frequency $I_\uparrow\approx 0$ (dashed line)
demonstrating the efficiency of the spin-polarized pump. For this
particular case, pumping of spin-down electrons occurs as one
electron with spin $\downarrow$ becomes delocalized (via a one
photon process) between both dots. If the chemical potential for
taking $\downarrow$ electrons out of the right dot fulfils
$\mu_{2,0}(1,2)=U_R-E_\uparrow>\mu_R$ while, on the other hand,
the chemical potential for taking $\uparrow$ electrons out of the
right dot fulfils $\mu_{2,1}(1,2)=U_R-E_\downarrow<\mu_R$ the
resulting pumped current is spin-down polarized. We emphasize
again that this pumping of spin polarized ($\downarrow$) electrons
is realized with unpolarized leads. Such spin-polarized current is
obtained either through the sequence
$(\downarrow\uparrow,\uparrow)\stackrel{\rm AC}\Leftrightarrow
(\uparrow,\downarrow\uparrow) \stackrel{\Gamma_R}\Rightarrow
(\uparrow,\uparrow)\stackrel{\Gamma_L}\Rightarrow
(\downarrow\uparrow,\uparrow)$ or, alternatively,
$(\downarrow\uparrow,\uparrow)\stackrel{\rm AC}\Leftrightarrow
(\uparrow,\downarrow\uparrow) \stackrel{\Gamma_L}\Rightarrow
(\downarrow\uparrow,\downarrow\uparrow)\stackrel{\Gamma_R}\Rightarrow
(\downarrow\uparrow,\uparrow)$. Reducing the frequency to
$\omega=\omega_{\downarrow}/2=0.15$, there is a second peak
(corresponding to the absorption of two photons) in the spin down
current. A three-photon process occurs at
$\omega=\omega_{\downarrow}/3=0.1$.

By increasing the frequency to $\omega=\omega_{\uparrow}\equiv
E_{(\downarrow, \uparrow
\uparrow)}-E_{(\downarrow\uparrow,\uparrow)}=
\Delta\epsilon+U_R-U_L-J/4=0.7$ (see Fig. 2), a current peak with
spin up polarization appears.
In this case, the pumped current occurs as one electron with spin
$\uparrow$ becomes delocalized between the states
$(\downarrow\uparrow, \uparrow)$ and $(\downarrow,
\uparrow\uparrow)$.
This spin up electron subsequently decays to the right reservoir
which leads to a pumped current through the sequence
$(\downarrow\uparrow,\uparrow)\stackrel{\rm AC}\Leftrightarrow
(\downarrow,\uparrow \uparrow)\stackrel{\Gamma_R}\Rightarrow
(\downarrow,\uparrow)\stackrel{\Gamma_L}\Rightarrow
(\downarrow\uparrow,\uparrow)$ or
$(\downarrow\uparrow,\uparrow)\stackrel{\rm AC}\Leftrightarrow
(\downarrow,\uparrow \uparrow)\stackrel{\Gamma_L}\Rightarrow
(\downarrow \uparrow,\uparrow
\uparrow)\stackrel{\Gamma_R}\Rightarrow
(\downarrow\uparrow,\uparrow)$. At $\omega_\uparrow$,
$I_\downarrow\approx 0$ such that the spin polarization, defined
as
\begin{equation}
P(\omega,V_{AC})\equiv
\frac{I_\uparrow-I_\downarrow}{I_\uparrow+I_\downarrow},
\end{equation}
{\it has been completely reversed by tuning the frequency of the
ac field}, namely
$P(\omega_\uparrow,V_{AC}^0)=1=-P(\omega_\downarrow,V_{AC}^0)$.
Note that the energy difference between both processes,
$\omega_\uparrow-\omega_\downarrow=\Delta\epsilon-J/4$ corresponds
to the energy difference between the triplet excited state and the
singlet ground state in the right dot {\it at zero magnetic
field}.

Reducing the frequency to $\omega=\omega_{\uparrow}/2$ and
$\omega=\omega_{\uparrow}/3$ and so on, peaks corresponding to
absorption of up to seven photons appear.
Note that each of these peaks has a different width. This
remarkable fact can be attributed to a renormalization of the
inter-dot hopping induced by the ac potential \cite{report1,
stafford}. At the frequency (or subharmonic) corresponding to the
energy difference between two levels, the Rabi frequency becomes
renormalized by the ac potential as $\Omega_{Rabi}$=$2t_{LR}
J_N(V_{AC}/\omega)$, $J_N$ is the Bessel function of order $N$
(=number of photons absorbed).
The width of the peaks is given by the coupling to the leads
provided that $\Gamma_{L,R}>\Omega_{Rabi}$. By contrast, if
$\Gamma_{L,R}<\Omega_{Rabi}$ the width of the current peak is
determined by $\Omega_{Rabi}$ . Since $\Omega_{Rabi}$ depends on
the number of photons in a nonlinear fashion (through the
dependence of $J_N$ on the ratio $V_{AC}/\omega$), it follows that
the widths of the peaks change in a non trivial way as a function
of $\omega$. A similar nonlinear dependence of the height of the
peaks as a function of the ratio $V_{AC}/\omega$ is expected. We
illustrate this nontrivial behavior in the inset of Fig. 2 where
we explore the low intensities regime ($V_{AC}=V_{AC}^0/5$=0.14).
In general, the trend
we obtain is consistent with previous analytical estimations
\cite{stafford}.

Another interesting feature of the spin pump is that there are
frequencies where the one-photon process corresponding to pumping
of $\downarrow$ electrons can overlap with multiphoton processes
corresponding to pumping of $\uparrow$ electrons. Thus, at these
frequencies the current is no longer fully spin-polarized.
One can use this to modify the polarization of the current by
changing $V_{AC}$ (at fixed $\omega$). We illustrate this with
Fig. 3, where the parameters are chosen such that the $N$=1 peak
of $I_\downarrow$ is centered at the same frequency
($\omega=\omega_\downarrow$= 0.3) as the $N$=2 peak of
$I_\uparrow$. At this frequency, the spin polarization can be
tuned by modifying the intensity of the ac potential (Fig.3,
inset). This result, together with those shown in Fig. 2,
demonstrate that the spin polarization of the current
$P(\omega,V_{AC})$ can be fully manipulated by tuning either the
frequency or the intensity of the external ac field.

\begin{figure}
\epsfig{file=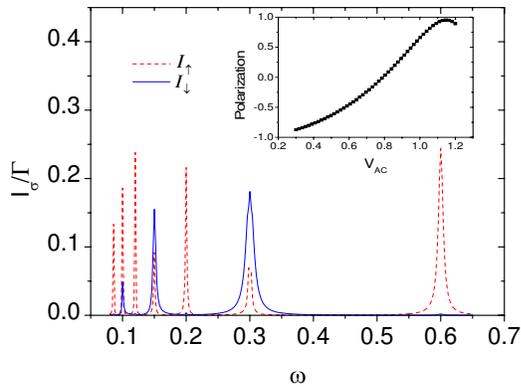,angle=0,width=0.450\textwidth}
\caption{(Color online) Pumped current as a function of the ac
frequency. Same parameters as Fig. 2 except $\Delta\epsilon=
0.35$. The interesting feature here is the overlap between the the
one-photon absorption peak of $I_\downarrow$ (solid line) and the
two-photon absorption peak of $I_\uparrow$ (dashed line) at
$\omega=\omega_\downarrow = 0.3$. The inset shows the spin
polarization $P(\omega,V_{AC})$ versus the ac intensity $V_{AC}$
for fixed frequency $\omega =\omega_\downarrow = 0.3$
demonstrating the possibility of controlling the spin polarization
of the current by tuning the intensity of the ac field.}
\end{figure}

Finally, it is important to note also that, contrary to the case
for spin-down pumping, the pumping of spin up electrons leaves the
double dot in the excited state $|\downarrow,\uparrow\rangle$.
This makes the spin-up current extremely sensitive to spin
relaxation processes. If the spin $\downarrow$ decays before the
next electron enters into the left dot, namely if
$W_{\uparrow\downarrow}\gtrsim\Gamma_L$, a spin-down current
appears through the cycle
$(\downarrow\uparrow,\uparrow)\stackrel{\rm AC}\Leftrightarrow
(\downarrow,\uparrow \uparrow)\stackrel{\Gamma_R}\Rightarrow
(\downarrow,\uparrow)\stackrel{W_{\uparrow\downarrow}}\Rightarrow
(\uparrow, \uparrow)\stackrel{\Gamma_L}\Rightarrow
(\uparrow\downarrow,\uparrow)$ and the pumping cycle is no longer
100\% spin-up polarized.
We study this effect in Fig. 4, where we plot the main PAT peak at
$\omega_{\uparrow}=0.6$ for increasing $W_{\uparrow\downarrow}$.
As one expects, the peak broadens as $W_{\uparrow\downarrow}$
increases. The full widths (FWHM) of the resonances
are plotted as a function of $W_{\uparrow\downarrow}$ in the
inset. For large intensities ($V_{AC}=V_{AC}^0$, circles) the
FWHM´s grow in a nonlinear fashion. This is reminiscent of the
so-called saturation regime, a well known phenomenon in the
context of optical Bloch equations \cite {boyd}. Note, however,
that there are three energy scales involved now in the dynamics of
the density matrix, $\Omega_{Rabi}$, $\Gamma$ and
$W_{\uparrow\downarrow}$, such that other sources of nonlinearity
(like the ones described when discussing Fig. 2) cannot be
completely ruled out \cite{nota}. In order to minimize nonlinear
effects we investigate the low intensity regime
($V_{AC}=V_{AC}^0/5$, squares) where we expect a FWHM dominated by
decoherence. The behavior is now linear with a slope which,
interestingly, approaches FWHM$\sim 20
W_{\uparrow\downarrow}=2/T_2$. Thus, experiments along these lines
would complement the information about decoherence extracted from
other setups \cite{Engel-Loss}.

\begin{figure}
\epsfig{file=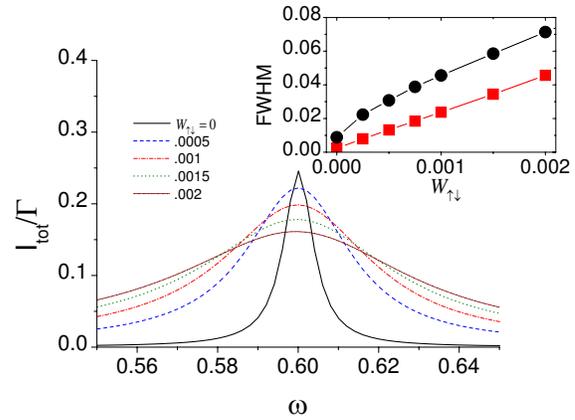,angle=0,width=0.450\textwidth}
\caption{(Color online) Pumped current near resonance
$\omega=\omega_\uparrow$= 0.6 for different relaxation rates.
Inset: FWHM of the total current as a function of relaxation rate,
for strong (black dots) and weak (red squares) field intensity}
\end{figure}

{\it Summary and experimental accessibility}.--In summary, we have
proposed and analyzed a new scheme of realizing {\it both} spin
filtering and spin pumping by using ac-driven double quantum dots
coupled to unpolarized leads. Our results demonstrate that the
spin polarization of the current can be manipulated (including
fully reversing) by just tuning the parameters of the ac field.
Our results also show that the width in frequency of the spin-up
pumped current gives information about spin decoherence in the
quantum dot. We finish by mentioning that our proposal is within
reach with today's technology for high-frequency experiments in
quantum dots \cite{report2,oosterkamp,petta}. Indeed, PAT with
two-electron spin states has recently been reported
\cite{wilfred-physicaE}.

We thank Wilfred van der Wiel for his help. Work supported by
Programa de Cooperaci\'on Bilateral CSIC-CONACYT, by grant
DGAPA-UNAM 114403-3 (E.C), by the EU grant HPRN-CT-2000-00144 and
by the Ministerio de Ciencia y Tecnolog\'{\i}a of Spain through
grant MAT2002-02465 (R. A. and G. P.) and the "Ram\'on y Cajal"
program (R. A.).


\begin{thebibliography}{11}
\bibitem{spintronics}
 S. A. Wolf {\it et al.},
 Science {\bf 294}, 1488 (2001).
\bibitem{ferro}
M. Johnson and R. H. Silsbee, Phys. Rev. Lett. {\bf 55}, 1790
(1985); F. J. Jedema {\it et al.},
Nature(London) {\bf 410}, 345 (2001).
\bibitem{semimag}
 R. Fiederling {\it et al.},
 Nature (London) {\bf 402}, 787
(1999);  Y. Ohno {\it et al.},
Nature (London) {\bf 402}, 790 (1999).
\bibitem{spin-pump1}
 E. R. Mucciolo {\it et al.},
 Phys. Rev. Lett. {\bf 89}, 146802 (2002); Susan K. Watson {\it et al.},
 Phys. Rev. Lett. {\bf 91}, 258301 (2003); M. G. Vavilov {\it et
 al.}, cond-mat/0410042.
\bibitem{spin-pump2}
T. Aono, Phys. Rev. B {\bf 67}, 155303 (2003); Qing-feng Sun {\it
et al.},
 Phys. Rev. Lett. {\bf 90}, 258301 (2003); E. Cota {\it et al.},
Nanotechnology {\bf 14}, 152-156 (2003).
\bibitem{hanson}
R. Hanson {\it et al.}
Phys. Rev. B, {\bf 70}, 241304 (2004).
\bibitem{recher}
P. Recher {\it et al.}
Phys. Rev. Lett., {\bf 85}, 1962 (2000).
\bibitem{Ono97-Johnson04}
K. Ono {\it et al.} Science {\bf 297}, 1313 (2002); A. C. Johnson
{\it et al.} cond-mat/0410679.
\bibitem{report1}
G. Platero and R. Aguado, Physics Reports, {\bf 395}, 1 (2004).
\bibitem{report2}
W. G. van der Wiel {\it et al.}, 'Photon assisted tunneling in
quantum dots', in: I.V. Lerner, et al. (Eds.), {\it Strongly
Correlated Fermions and Bosons in Low-dimensional Disordered
Systems}, Kluwer Academic Publishers, pp. 43-68 (2002).
\bibitem{stafford}
C. A. Stafford and N. S. Wingreen Phys. Rev. Lett. {\bf 76}, 1916
(1996).
\bibitem{hazelzet}
B. L. Hazelzet, M. R. Wegewijs, T. H. Stoof, and Yu. V. Nazarov,
Phys. Rev. B {\bf 63}, 165313 (2001).
\bibitem{oosterkamp}
T. H. Oosterkamp {\it et al.}, Nature (London) {\bf 395}, 873
(1998).
\bibitem{petta}
J. R. Petta {\it et al.}, cond-mat/0408139.
\bibitem{hanson2}
R. Hanson {\it et al.},
Phys. Rev. Lett. {\bf 91}, 196802 (2003).
\bibitem{Loss-relaxation}
A detailed study of spin relaxation and decoherence in a GaAs
quantum dot due to spin-orbit interaction can be found in, Vitaly
N. Golovach {\it et al.},
Phys. Rev. Lett. {\bf 93}, 016601 (2004).
\bibitem{boyd}
Robert W. Boyd, {\it Non-linear Optics} (Academic Press, NY 2003)
\bibitem{nota}
A detailed analytical study of the dynamics of the effective few
level problem (to be published elsewhere) is needed in order to
substantiate these arguments.
\bibitem{Engel-Loss}
See also, Hans-Andreas Engel and Daniel Loss, Phys. Rev. Lett.,
{\bf 86}, 4648 (2001).
\bibitem{wilfred-physicaE}
T. Kodera, W. G. van der Wiel, K. Ono, S. Sasaki, T. Fujisawa, and
S. Tarucha, Physica E, {\bf 22}, 518 (2004).
\end{thebibliography}
\end{document}